# SPARSE-VIEW CT RECONSTRUCTION VIA CONVOLUTIONAL SPARSE CODING

*Peng Bao[1], Wenjun Xia[1], Kang Yang[1], Jiliu Zhou[1], Senior Member, IEEE and Yi Zhang[1, *], Senior Member, IEEE*

[1]College of Computer Science, Sichuan University, Chengdu 610065, China

## ABSTRACT

Traditional dictionary learning based CT reconstruction methods are patch-based and the features learned with these methods often contain shifted versions of the same features. To deal with these problems, the convolutional sparse coding (CSC) has been proposed and introduced into various applications. In this paper, inspired by the successful applications of CSC in the field of signal processing, we propose a novel sparse-view CT reconstruction method based on CSC with gradient regularization on feature maps. By directly working on whole image, which need not to divide the image into overlapped patches like dictionary learning based methods, the proposed method can maintain more details and avoid the artifacts caused by patch aggregation. Experimental results demonstrate that the proposed method has better performance than several existing algorithms in both qualitative and quantitative aspects.

## 1. INTRODUCTION

In recent years, reconstructing high quality images from undersampled projection data is an attractive research topic in the field of CT imaging. Traditional analytic algorithms, such as FBP, cannot reconstruct high quality images with insufficient projection data. On the other side, iterative reconstruction algorithms are effective under this situation, which can guarantee the image quality when the projection views are incomplete. However, when projection views are highly sparse, it is difficult to achieve a satisfactory result without extra prior information. Introducing reasonable priors can significantly improve imaging quality of traditional iterative methods, such as algebraic reconstruction technique (ART) and expectation maximization (EM). With the development of compressed sensing (CS) theory, Sidky et al. [1] coupled total variation (TV) and projection onto convex sets (POCS) to solve incomplete projection data reconstruction problem and achieved good performance. Following this path, many variants of TV have been proposed. Particularly, Niu *et al*. [2] proposed to combine penalized weighted least-squares (PWLS) [3] with total generalized variation (TGV) [4] to reduce the undesired patchy effects caused by the mathematical assumption of TV. However, these methods, including PWLS-TGV, still suffer from patchy effects to different degrees.

As another representative works, dictionary learning (DL) methods [5] have been proved effective in CT reconstruction in [6]. Two kinds of dictionaries are learned as bases: (1) a global dictionary can be learned from an external training set that is fixed during the reconstruction process, and (2) an adaptive dictionary learned from the intermediate result that keeps updating during the iterations.

However, most dictionary learning methods are patch-based and the learned features often contain shifted versions of the same features. To address these problems, convolutional sparse coding (CSC) has been proposed in which shift invariance is directly modelled in the objective function. CSC has been demonstrated very useful in a wide range of computer vision problems. Aided by $l_2$ penalty on the gradient of the feature maps, CSC can be further improved for impulse noise denoising problem [7]. Inspired by this successful research, in this work, we proposed a penalized weighted least-squares method based on CSC aided by gradient regularization on feature maps for sparse-view CT reconstruction (PWLS-CSCGR). The remainder of this paper is organized as follows. In Section 2, the proposed reconstruction scheme will be elaborated. In Section 3, experimental designs and representative results are given. Finally, we will conclude this paper in the Section 4.

## 2. METHOD

### 2.1. Penalized Weighted Least-Squares

The calibrated and log-transformed projection data approximately follow a Gaussian distribution, which can be described by the following formula:

$$\sigma_p^2 = r_p \times exp(\overline{y}_p/\varepsilon) \quad (1)$$

where $\overline{y}_p$ and $\sigma_p^2$ are the mean and variance of the measured projection data at $p$-th bin respectively, $r_p$ is a parameter adpative to different bins and $\varepsilon$ is a scaling parameter. Based on these properties of projection data, the PWLS CT image reconstruction can be modelled as follows:

$$arg\,min_u (y - Au)^T \Sigma^{-1}(y - Au) + \beta R(u) \quad (2)$$

where $y$ denotes the projection data, $u$ is the vectorized attenuation coefficients to be reconstructed and $(\cdot)^T$ denotes the transpose operation. $A$ is the system matrix with size of $P \times Q$ ($M$ is the total number of projection data and $Q$ is the total number of image pixels). $\Sigma$ is a diagonal matrix with the $p$-th element of $\sigma_p^2$ as calculated by (1). $R(u)$ represents the regularization term and $\beta$ is a regularization parameter.

### 2.2. PWLS-CSCGR

Current TV and DL based regularization methods suffer from blocky effects or patch aggregation artifacts. To circumvent these problems, in this paper, we propose to utilize CSC with gradient regularization on feature maps as the regularization term in (2). Then the following minimization model is obtained:

$$arg\ min_{u,\{M_i\},\{f_i\}}\ \frac{1}{2}(y-Au)^T\Sigma^{-1}(y-Au) +$$
$$\beta(\frac{1}{2}\left\|\sum_{i=1}^{N}f_i * M_i - u\right\|_2^2 + \lambda\sum_{i=1}^{N}\|M_i\|_1 +$$
$$\frac{\tau}{2}\sum_{i=1}^{N}\left\|\sqrt{(g_0 * M_i)^2 + (g_1 * M_i)^2}\right\|_2^2). \quad (3)$$

where $*$ is convolution operator, $\{f_i\}$ is a set of filters, $M_i$ is the feature map corresponding to filter $f_i$, $g_0$ and $g_1$ are the filters that compute the gradient along image rows and columns respectively, and $\lambda$ and $\tau$ are the hyper-parameters. In this model, we use predetermined filters $\{f_i\}$, which can be trained by method proposed in [8]. Then the model becomes following optimization problem:

$$arg\ min_{u,\{M_i\}}\ \frac{1}{2}(y-Au)^T\Sigma^{-1}(y-Au) +$$
$$\beta(\frac{1}{2}\left\|\sum_{i=1}^{N}f_i * M_i - u\right\|_2^2 + \lambda\sum_{i=1}^{N}\|M_i\|_1 +$$
$$\frac{\tau}{2}\sum_{i=1}^{N}\left\|\sqrt{(g_0 * M_i)^2 + (g_1 * M_i)^2}\right\|_2^2). \quad (4)$$

An alternating minimization scheme can be applied to solve (4). First, an intermediate reconstructed image $u$ can obtained with a set of fixed feature maps $\{\widetilde{M}_i\}$. Thus, (4) is transformed to:

$$arg\ min_u\ (y-Au)^T\Sigma^{-1}(y-Au) +$$
$$\beta\left\|\sum_{i=1}^{N}f_i * \widetilde{M}_i - u\right\|_2^2. \quad (5)$$

With the separable paraboloid surrogate method [6], the solution of (5) can be obtained, which is expressed as follows:

$$u_q^{t+1} = u_q^t - \frac{\sum_{p=1}^{P}\left((1/\sigma_p^2)a_{pq}\left([Au^t]_p - y_p\right)\right)}{\sum_{p=1}^{P}\left((1/\sigma_p^2)a_{pq}\sum_{k=1}^{Q}a_{pk}\right) + \beta}$$
$$-\frac{\beta\left(\left[\sum_{i=1}^{N}f_i * \widetilde{M}_i\right]_q^t - u_q^t\right)}{\sum_{p=1}^{P}\left((1/\sigma_p^2)a_{pq}\sum_{k=1}^{Q}a_{pk}\right) + \beta},$$
$$q = 1,2,\dots,Q, \quad (6)$$

where $t = 0,1,2,\dots,T$ represents the iteration index and $a_{pq}$ is the element of $A$. The second step is to re-express the intermediate result $u$ with the fixed filters $\{f_i\}$. Since CSC does not provide a good representation for low-frequency component, only high-frequency component of $u$ is expressed by CSC. Then we have the following optimization problem:

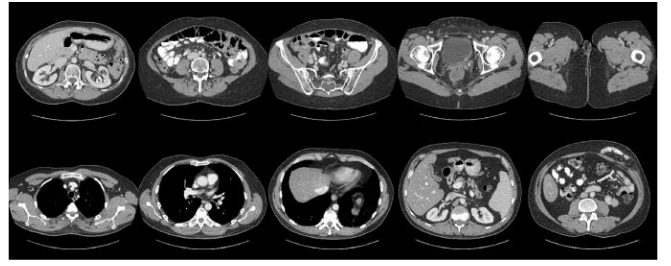

Fig. 1. The images in the training set. The display window is [-150 250] HU.

$$arg\ min_{\{M_i\}}\ \frac{1}{2}\left\|\sum_{i=1}^{N}f_i * M_i - u\right\|_2^2 + \lambda\sum_{i=1}^{N}\|M_i\|_1 +$$
$$\frac{\tau}{2}\sum_{i=1}^{N}\left\|\sqrt{(g_0 * M_i)^2 + (g_1 * M_i)^2}\right\|_2^2, \quad (7)$$

Define the linear operators $G_0$ and $G_1$ such that $G_l M_i = g_l * M_i$, then the gradient term in (7) can be rewritten as

$$\frac{\tau}{2}\sum_{i=1}^{N}\|G_0 M_i\|_2^2 + \frac{\tau}{2}\sum_{i=1}^{N}\|G_1 M_i\|_2^2. \quad (8)$$

If we define $F_i M_i = f_i * M_i$, $F = (F_1\ F_2\ \cdots\ F_N)$, $M = (M_1\ M_2\ \cdots\ M_N)^T$, and

$$\Phi_l = \begin{pmatrix} G_l & 0 & \cdots \\ 0 & G_l & \cdots \\ \vdots & \vdots & \ddots \end{pmatrix}, \quad (9)$$

(7) can be reformulated as follows:
$$arg\ min_M\ \frac{1}{2}\|FM - u\|_2^2 + \lambda\|M\|_1 + \frac{\tau}{2}\|\Phi_0 M\|_2^2 + \frac{\tau}{2}\|\Phi_1 M\|_2^2, \quad (10)$$

Alternating Direction Method of Multipliers (ADMM) is applied to solve (10) by introducing $B$ that is constrained to be equal to the variable $M$, leading to following problem:

$$arg\ min_{M,B}\ \frac{1}{2}\|FM - u\|_2^2 + \frac{\tau}{2}\|\Phi_0 M\|_2^2 + \frac{\tau}{2}\|\Phi_1 M\|_2^2 + \lambda\|B\|_1,\ s.t.\ M = B. \quad (11)$$

With ADMM, we have iterations as follows:
$$M^{j+1} = arg\ min_M \frac{1}{2}\|FM - u\|_2^2 + \frac{\tau}{2}\|\Phi_0 M\|_2^2 + \frac{\tau}{2}\|\Phi_1 M\|_2^2 + \frac{\rho}{2}\|M - B^j + C^j\|_2^2 \quad (12)$$
$$B^{j+1} = arg\ min_B \lambda\|B\|_1 + \frac{\rho}{2}\|M^{j+1} - B + C^j\|_2^2 \quad (13)$$
$$C^{j+1} = C^j + M^{j+1} - B^{j+1} \quad (14)$$

(12) can be transformed into Fourier domain and the solution can be expressed as:
$$(\widehat{F}^H\widehat{F} + \tau\widehat{\Phi}_0^H\widehat{\Phi}_0 + \tau\widehat{\Phi}_1^H\widehat{\Phi}_1 + \rho I)\widehat{M}$$
$$= \widehat{F}^H\widehat{u} + \rho(\widehat{B} - \widehat{C}), \quad (15)$$

where $\widehat{F}$, $\widehat{\Phi}_0$, $\widehat{\Phi}_1$, $\widehat{M}$, $\widehat{u}$, $\widehat{B}$ and $\widehat{C}$ denote the expressions after transforming into Fourier domain. $I$ is the identity matrix. (15) can be solved efficiently by exploiting the Sherman Morrison formula [8]. And the closed-form solution for (13) can be obtained by:

$$B^{j+1} = S_{\lambda/\rho}(M^{j+1} + C^j), \quad (16)$$

where $S(\cdot)$ denotes the soft-thresholding function.

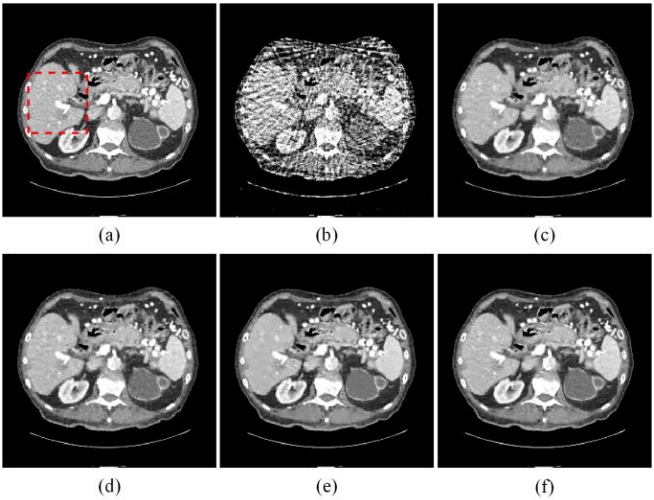

Fig. 2. Abdominal images reconstructed by various methods. (a) The reference image versus the images reconstructed by (b) FBP, (c) TV-POCS, (d) PWLS-TGV, (e) PWLS-DL and (f) PWLS-CSCGR. The display window is [-150 250] HU.

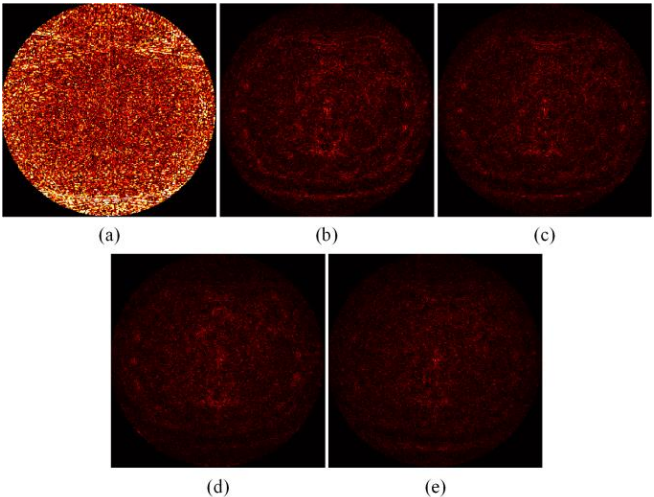

Fig. 3. Difference images relative to the original image. (a) FBP, (b) TV-POCS, (c) PWLS-TGV, (d) PWLS-DL and (e) PWLS-CSCGR. The display window is [-1000 -700] HU.

Table 1: Quantitative results obtained by different algorithms for the abdominal and thoracic images

|  | ABDOMINAL | | | THORACIC | | |
|---|---|---|---|---|---|---|
|  | PSNR | RMSE | SSIM | PSNR | RMSE | SSIM |
| FBP | 25.35 | 0.05417 | 0.45858 | 22.99 | 0.07087 | 0.42335 |
| TV-POCS | 43.04 | 0.00704 | 0.97187 | 42.92 | 0.00714 | 0.97751 |
| PWLS-TGV | 44.05 | 0.00627 | 0.97697 | 44.43 | 0.00600 | 0.98310 |
| PWLS-DL | 44.84 | 0.00573 | 0.97975 | 43.68 | 0.00655 | 0.98145 |
| PWLS-CSCGR | **45.73** | **0.00517** | **0.98329** | **46.07** | **0.00497** | **0.98753** |

## 3. EXPERIMENTAL RESULTS

In this section, some experiments were tested to validate the performance of the proposed method. All the images authorized by Mayo Clinic and downloaded from "the NIH-AAPM-Mayo Clinic Low Dose CT Grand Challenge". 10 images were randomly selected as training set to train the predetermined filters $\{f_i\}$. Fig. 1 shows the images in the training set. The geometry configuration of the simulations was set identical to [9]. 64 projection views were evenly distributed over 360°. Several methods were compared with our pro-posed algorithm, including FBP, TV-POCS [1], PWLS-TGV [2] and PWLS-DL [6]. Specifically, for PWLS-DL, the number of atoms in dictionary was set to 256 and the number of nonzero elements in each coefficient vector was set to 16. The parameters of the proposed method were experimentally set as follows: filter size was set to $10 \times 10$, the number of filters was 32, $\beta = \lambda = 0.005$, $\rho = 100 \times \lambda + 1$ and $\tau = 0.06$. All the parameters had same initializations in all the experiments to demonstrate the robustness of our method.

### 3.1. Abdominal Case

The original abdominal image and the results reconstructed from different methods are shown in Fig. 2. The result of FBP contains severe streak artifacts and is almost clinically useless. All the other algorithms remove the artifacts efficiently. However, in Fig. 2(c) and (d), blocky effects are noticed to different degrees. PWLS-DL effectively avoided the blocky effects in Fig. 2(e), but the result looks smooth in some organs. The reason for this should rely on that the DL based method is patch-based, which will average all the overlapped patches to produce the final result. This procedure may suppress the noise efficiently, but some small details may be smoothened too. In Fig. 2(f), the proposed method maintains more details than all the other methods. The absolute difference images are illustrated in Fig. 3 and it can be observed that result reconstructed by our method is closest to the reference image.

To further demonstrate the details, Fig. 4 shows the results of magnified region of interest (ROI), which is indicated by the red box in Fig. 2(a). The red arrows indicate several contrast enhanced blood vessels in the liver, which can only be well identified by our method. Specially, in Fig. 4(e) PWLS-DL smoothens the liver region, and the contrast enhanced blood vessels are lost. In Fig. 4(f), the proposed method preserves most details and has most coherent visual effect to the reference image, even for the mottle-like structures in the liver. The quantitative results are given in Table 1. The proposed algorithm outperforms other methods for all the metrics.

### 3.2. Thoracic Simulation

Fig. 5 presents the thoracic images reconstructed by different methods. The FBP result is covered by the artifacts and it is hard to discriminate the artifacts and blood vessels in the lungs. In Fig. 5(c) - (e), most artifacts are suppressed, but the blood vessels in some area, which are indicated by the red arrows, are blurred. In Fig. 5(f), more details in the lungs are preserved while the artifacts are well eliminated. Particularly, a region which is indicated by the red rectangle in Fig. 5(a) is enlarged in Fig. 6. It can be observed that the curve-like structures indicated by the red arrow are clearly visible in the result of proposed method and all the other methods cannot maintain these details. The quantitative evaluation of the results of thoracic image is given in the right part of Table 1. It is consistent to the results of abdominal image that our method still had the best scores.

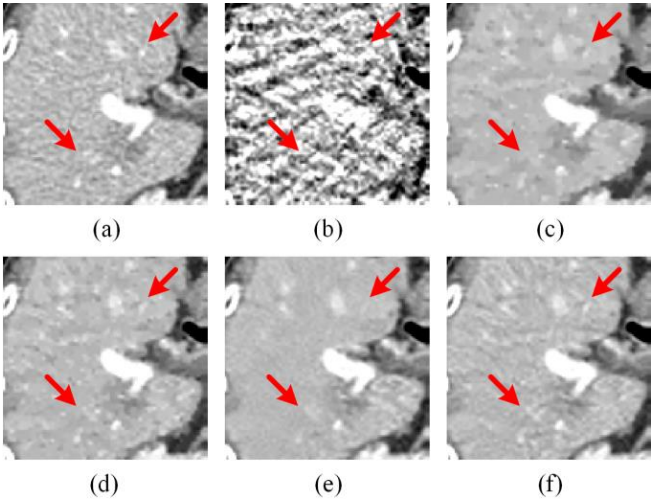

Fig. 4. Zoomed ROI indicated by the red rectangle in 2(a). (a) The reference image versus the images reconstructed by (b) FBP, (c) TV-POCS, (d) PWLS-TGV, (e) PWLS-DL and (f) PWLS-CSCGR. The display window is [-150 250] HU.

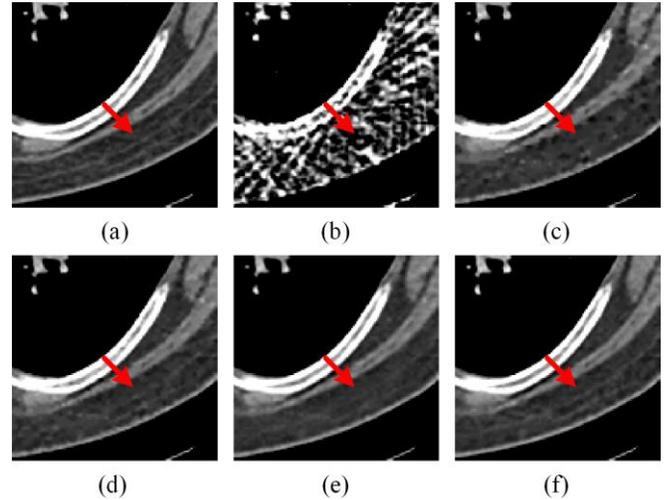

Fig. 6. Zoomed ROI indicated by the red rectangle in 5(a). (a) The reference image versus the images reconstructed by (b) FBP, (c) TV-POCS, (d) PWLS-TGV, (e) PWLS-DL and (f) PWLS-CSCGR. The display window is [-150 250]HU.

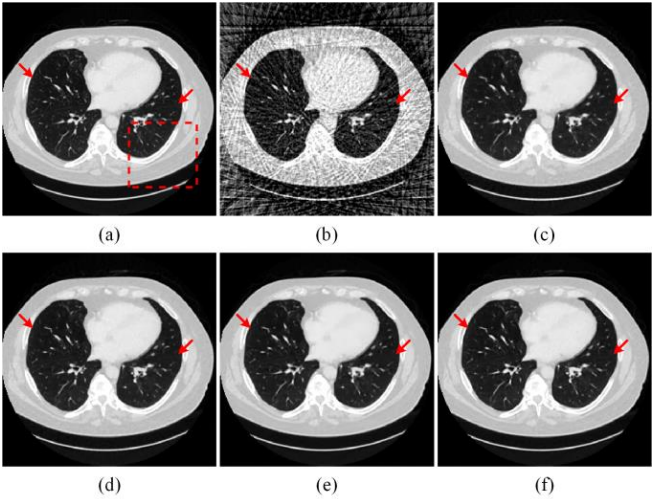

Fig. 5. Thoracic images reconstructed by various methods. (a) The reference image versus the images reconstructed by (b) FBP, (c) TV-POCS, (d) PWLS-TGV, (e) PWLS-DL and (f) PWLS-CSCGR. The display window is [-1000 250] HU.

## 4. CONCLUSION

In this paper, we proposed a novel sparse-view CT reconstruction algorithm, combining PWLS and CSCGR. The proposed method directly performed on whole image rather than overlapped patches, which can avoid the artifacts caused by patch aggregation and preserve more details. Several experiments demonstrated that the proposed method had better performance than competing methods. In the future, we will try to extend this method to other topics, such as metal artifact reduction. We can also optimize our model with deep learning based methods [10, 11].

### ACKNOWLEDGMENT

This work was supported in part by the National Natural Science Foundation of China under grants 61671312, the Sichuan Science and Technology Program under grants 2018HH0070 and Miaozi Project in Science and Technology Innovation Program of Sichuan Province under grants 18-YCG041.